\documentclass[aps,preprint,epsfig]{revtex4}
\usepackage{graphicx,latexsym,amsmath,float,caption}
\begin{document}

\title{First Passage of a Randomly Accelerated Particle}

\author{Theodore W. Burkhardt}

\affiliation{Department of Physics, Temple University,\\
Philadelphia, PA 19122, U.S.A. \\
tburk@temple.edu}

\begin{abstract}
In the random acceleration process, a point particle is accelerated according to $\ddot{x}=\eta(t)$, where the right hand side represents Gaussian white noise with zero mean. We begin with the case of a particle with initial position $x_0$ and initial velocity $v_0$ and review the statistics of its first arrival at the origin and its first return to the origin. Multiple returns to the origin, motion with a constant force in addition to a random force, and persistence properties for several boundary conditions at the origin are also considered. Next we review first-exit properties of a randomly accelerated particle from the finite interval $0<x<1$. Then the close connection between the extreme value statistics of a randomly accelerated particle and its first-passage properties is discussed. Finally some applications where first-passage statistics of the random acceleration process play a role are considered.
\end{abstract}
\maketitle

\section{Introduction}\label{intro}
For a particle moving in one dimension subject to a random force in the form
of Gaussian white noise, the Newtonian equation of motion has the form
\begin{equation}
\ddot{x}=\eta(t),\quad \langle\eta(t)\rangle=0,\quad \langle\eta(t)\eta(t')\rangle=
\Lambda^{-1}\delta(t-t').\label{eqmo}
\end{equation}
This is the random acceleration process considered in this chapter.
For a given $\eta(t)$, the position $x$ and velocity $v$ of the particle evolve from the initial values $x_0$ and $v_0$ at $t=0$ according to
\begin{eqnarray}
&&v(t)=v_0+\int_0^t\eta(t')\thinspace dt',\label{v(t)}\\
&&x(t)=x_0+\int_0^t v(t')\thinspace dt'=x_0+v_0t+\int_0^t(t-t')\eta(t')\thinspace dt'.\label{x(t)}
\label{x(t)}
\end{eqnarray}
Thus, $v(t)$ corresponds to a Brownian curve or random walk, and $x(t)$ to the integral of a Brownian curve.

The propagator or probability density $P(x,v;x_0,v_0;t)$ for propagation from the initial values $x_0,v_0$ to $x,v$ in a time $t$ will play a central role in our discussion of first-passage properties. It has the path integral representation
\begin{equation}
P(x,v;x_0,v_0;t)=\int Dx\exp\left[-{\Lambda\over 2}\int_0^t\left({d^2x\over dt^2}
\right)^2 dt\right]\label{pathintegral}
\end{equation}
and satisfies the Fokker-Planck equation \cite{hr}
\begin{equation}
\left({\partial\over\partial t}+v{\partial\over\partial x}-{1\over
2\Lambda}\;{\partial^2\over\partial v^2}\right)P(x,v;x_0,v_0;t) =0\label{fpeq}
\end{equation}
with initial condition
\begin{equation}
P(x,v;x_0,v_0;0)=\delta(x-x_0)\delta(v-v_0).\label{initcond}
\end{equation}
The factor $(2\Lambda)^{-1}$ in Eq. (\ref{fpeq}) can be eliminated by rescaling
$t$, and except in Sec. \ref{applications}, where applications to
semiflexible polymers are considered and $\Lambda$ corresponds to the persistence length, we set $(2\Lambda)^{-1}=1$ throughout this chapter.

In the unbounded geometry $-\infty<x<\infty$, Eqs. (\ref{fpeq}) and (\ref{initcond})
have the solution \cite{hr}
\begin{eqnarray}
&&P_0(x,v;x_0,v_0;t)=3^{1/2}(2\pi)^{-1}t^{-2}\nonumber\\
&&\quad\times\exp\left\{-3t^{-3}\left[(x-x_0-vt)(x-x_0-v_0t)+{1\over 3}(v-v_0)^2
t^2\right]\right\},\label{P0}
\end{eqnarray}
which is normalized so that
\begin{equation}
 \int_{-\infty}^\infty dx\int_{-\infty}^\infty dv\;P_0(x,v;x_0,v_0;t)=1.\label{norm}
\end{equation}

In this chapter we review theoretical work on the first-passage properties of random acceleration and also consider some applications where these properties play a central role. For a related review of the random acceleration process in bounded geometries, but with less focus on first-passage properties, see Ref. \cite{twb07}. The random acceleration process is also considered in the  reviews \cite{sr,bms} of first-passage properties of a wide variety of processes.

In Sec. \ref{halfline} we consider the first arrival at the origin of a randomly accelerated particle which is initially released at point $x_0$ with velocity $v_0$, the first return to the origin, and multiple returns to the origin. The pioneering contributions of McKean \cite{mck} and Marshall and Watson \cite{mw} are reviewed. The results are generalized to include a constant force, such as gravity, in addition to the random force. First-passage properties follow from the solution of the Fokker-Planck equation on the half line $x>0$ with an absorbing boundary condition at the origin. Some results for partially absorbing and inelastic boundaries are also briefly discussed.

In Sec. \ref{finiteinterval} the initial position $x_0$ of the randomly accelerated particle is assumed to lie in the finite interval $0<x<1$. Theoretical results for the mean exit time of a randomly accelerated particle from the interval and the probabilities that the first exit takes place at $x=0$ and at $x=1$ are reviewed.

Sec. \ref{extreme} considers the extreme value statistics of the random acceleration process and its close relation to first-passage statistics.

In Sec. \ref{applications} several mathematical and physical applications which can be analyzed in terms of first-passage properties of the random acceleration process are discussed or referenced. For example, the partition function of a long, semi-flexible polymer chain, fluctuating in equilibrium in a narrow cylindrical channel, is closely related to the first-exit probability of a randomly accelerated particle from a finite domain.

\section{Random Acceleration on the Half Line $x>0$}\label{halfline}
\subsection{First arrival at the origin}\label{firstarrival}
   We begin by considering the probability density or half-line propagator  $P(x,v;x_0,v_0;t)$
for propagation of the particle from $x_0,v_0$ to $x,v$ in a time $t$ {\em without leaving the positive $x$ axis}. This quantity satisfies the Fokker-Planck equation (\ref{fpeq}) with initial condition (\ref{initcond}) and boundary condition
\begin{equation}
 P(0,v;x_0,v_0;t)=0,\quad v>0.\label{absorbbc}
\end{equation}
 This boundary condition rightfully excludes all trajectories with $x=0$, $v>0$ at time $t$ from the propagator, since the corresponding particle has not remained on the positive real axis for a time $t$ but is returning after leaving it an earlier time. Following common usage, we refer to Eq. (\ref{absorbbc}) as the``absorbing" boundary condition, since it obviously applies if the particle sticks permanently at the origin on arrival. In contrast to Eq. (\ref{absorbbc}), $P(0,v;x_0,v_0;t)$ does not vanish for $v<0$ and is directly related, as indicated in the next paragraph, to the statistics of arrival at the origin. For $v<0$, $P(0,v;x_0,v_0;t)$ is not known at the outset, but is determined by solving the Fokker-Planck equation (\ref{fpeq}) with initial condition (\ref{initcond}) and boundary condition (\ref{absorbbc}), as constitutes a ``well-posed" problem \cite{mw}.

The ``survival probability" or probability $Q(x_0,v_0;t)$ that the randomly accelerated particle has not yet left the positive $x$ axis in a time $t$ is related to the half-line propagator by
\begin{equation}
 Q(x_0,v_0;t)=\int_{-\infty}^\infty dv\int_0^\infty dx\;P(x,v;x_0,v_0;t)
 \label{Qdef}
\end{equation}
and, according to Eq. (\ref{fpeq}) and the time reversal property $P(x,v;x_0,v_0;t)=P(x_0,-v_0;x,-v;t)$, satisfies the backward Fokker-Planck equation
\begin{equation}
\left({\partial\over\partial t}-v_0\thinspace{\partial\over\partial x_0}-\;{\partial^2\over\partial v_0^2}\right)Q(x_0,v_0;t) =0,\label{fpeqQ}
\end{equation}
with the initial condition and absorbing boundary condition
\begin{eqnarray}
&&Q(x_0,v_0;0)=1,\quad x_0>0,\label{initcondQ}\\
&&Q(0,v_0;t)=0,\quad v_0<0,\label{absorbbcQ}
\end{eqnarray}
respectively. Equations (\ref{fpeq}), (\ref{absorbbc}), and (\ref{Qdef}) imply
\begin{equation}
 {\partial\over\partial t}\thinspace Q(x_0,v_0;t)
 =- \int_0^\infty dv\;vP(0,-v;x_0,v_0;t).\label{probcons}
 \end{equation}
Thus, the probability
that the randomly accelerated particle leaves the the positive $x$ axis for the first time between $t$ and $t+dt$ and with speed between $v$ and $v+dv$, is given by $vP(0,-v;x_0,v_0;t)\thinspace dv\thinspace dt$,
where $v>0$.

In the half space $(x,v)$ with $x>0$, the propagator $P(x,v;x_0,v_0;t)$ vanishes identically on the positive $v$ axis, according to Eq. (\ref{absorbbc}), but otherwise is
positive. Solving the Fokker-Planck equation (\ref{fpeq}) with this boundary
condition is a formidable task. The equation has separable solutions
of the form $e^{st-Fx}{\rm Ai}(-F^{1/3}v+F^{-2/3}s)$, where ${\rm Ai(z)}$ is the
standard Airy function \cite{as}, but these solutions do not satisfy Eq.
(\ref{absorbbc}). The problem of superposing the separable solutions in order to
satisfy the boundary condition was solved by Marshall and Watson [{\cite{mw} in 1985, as discussed in Sec. \ref{marshallwatson}.

\subsection{First return to the origin}\label{mckean}

Two decades before the work of Marshall and Watson, the first-return distribution $vP_0(0,-v;0,v_0;t)$ of a particle that that begins at the origin with $v_0>0$ was already known from the ground-breaking work of McKean \cite{mck}. In 1963 he calculated this quantity, but not by solving the Fokker-Planck equation. A randomly accelerated particle which begins at the origin at $t=0$ with a positive velocity $v_0$ and crosses the origin at a later time $t$ with a positive velocity $v$ must have left the positive $x$ axis for the the first time at some earlier time $t_1$. This obvious statement is quantified in the integral equation
\begin{eqnarray}
&&P_0(0,v;0,v_0;t)\nonumber\\ &&\quad =\int_0^{t} dt_1\int_0^\infty dv_1\thinspace P_0(0,v;0,-v_1;t-t_1)\thinspace v_1
P(0,-v_1;0,v_0;t_1)\label{mck1}
\end{eqnarray}
for positive $v$ and $v_0$, which relates the free propagator $P_0$ given in Eq. (\ref{P0}) and the half-line propagator $P$ for the absorbing boundary condition (\ref{absorbbc}).

Performing Laplace and Bessel transformations $t\to s$ and $v\to\gamma$, respectively, McKean converted integral equation (\ref{mck1}) to the form
\begin{eqnarray}
&&\left[\thinspace 2v_0\cosh(\pi\gamma/3)\thinspace\right]^{-1}K_{i\gamma}\left(\sqrt{4s}\thinspace v_0\right)
\nonumber\\
&&\quad =\int_0^\infty dv_1\thinspace K_{i\gamma}\left(\sqrt{4s}\thinspace v_1\right)\tilde{P}(0,-v_1;0,v_0;s)\label{mck2}
\end{eqnarray}
for arbitrary positive $\gamma$, where $K_\nu(u)$ is the standard modified Bessel function \cite{as} and $\tilde{P}(x,v;x_0,v_0;s)=\int_0^\infty dt\;e^{-st}P(x,v;x_0,v_0;t)$.
Solving Eq. (\ref {mck2}) with the help of the Kontorovitch-Lebedev transformation \cite{kl},
\begin{eqnarray}
&&f(u)={2\over\pi^2}\int_0^\infty d\gamma\thinspace\gamma\sinh(\pi\gamma) g(\gamma)K_{i\gamma}(u),\label{kl1}\\
&&g(\gamma)=\int_0^\infty du\thinspace u^{-1}f(u)K_{i\gamma}(u),\label{kl2}
\end{eqnarray}
and performing the inverse Laplace transformation $s\to t$, he obtained
\begin{eqnarray}
&&\tilde{P}(0,-v;0,v_0;s)\nonumber\\ &&\qquad={2\over\pi^2v v_0}\int_0^\infty d\gamma\thinspace{\gamma\sinh(\pi\gamma)\over 2\cosh(\pi\gamma/3)}\thinspace K_{i\gamma}\left(\sqrt{4s}\thinspace v\right)K_{i\gamma}\left(\sqrt{4s}\thinspace v_0\right)\label{mck2}
\end{eqnarray}
and
\begin{eqnarray}
&&vP(0,-v;0,v_0;t)\nonumber\\ &&\qquad={\sqrt 3\over 2\pi}\;{v\over t^2} \exp\left[-\left(v^2-v
v_0+v_0^2\right)/t\right] {\rm erf}\left[\left(3v
v_0/t\right)^{1/2}\right]\label{mck3}
\end{eqnarray}
for the first-return distribution. Here $v$ and $v_0$ are both positive, and ${\rm
erf}(z)$ is the error function \cite{as}. 

In working with the integrals in this Section and Sec. \ref{lachal} involving integration of Bessel functions $K_{i\gamma}(z)$  over $\gamma$, such as Eq. (\ref{mck2}), it is useful to keep the asymptotic form
 \begin{equation}
 K_{i\gamma}(z)\approx \left({2\pi\over\gamma}\right)^{1/2}e^{-\pi\gamma/2}
 \cos\left(\gamma\ln{2\gamma\over z}-\gamma -{\pi\over 4}\right)
 \end{equation}
 for $\gamma\to\infty$ with $z$ fixed in mind. For completeness we note that, in contrast to Eq. (\ref{mck2}), $\tilde{P}(0,-v;0,v_0;s)$ may be written without special functions in the form
\begin{eqnarray}
&&\tilde{P}(0,-v;0,v_0;s)\nonumber\\ &&\ \ ={3\over 2\pi}\thinspace\left(vv_0\right)^{1/2}\int_0^1 dy\thinspace \exp\left[-2s^{1/2}\left(v^2-vv_0+v_0^2+3vv_0y^2\right)^{1/2}\right]\nonumber\\
&&\ \ \times\left[\left(v^2-vv_0+v_0^2+3vv_0y^2\right)^{-3/2}\right.\nonumber\\
&&\left.+
2s^{1/2}\left(v^2-vv_0+v_0^2+3vv_0y^2\right)^{-1}\right].\label{mck4}
\end{eqnarray}
This follows follows from Laplace tranformation of Eq. (\ref{mck3}) after substitution of a standard integral representation of the error function \cite{as}.

Substituting Eq. (\ref{mck3}) into Eq. (\ref{probcons}) and integrating over $v$ yields
\begin{eqnarray}
 &&-{\partial\over\partial t}\thinspace Q(0,v_0;t)=I_1(v_0,t)\nonumber\\&&\quad={3\over 2\pi^{3/2}}\thinspace v_0^{-2}e^{-v_0^2/t}\sum_{n=0}^\infty{\Gamma({5\over 4}+{n\over 2})\over n!}\thinspace\thinspace_2F_1(\textstyle{{1\over 2},-n;{3\over 2};3})\left(\displaystyle{v_0^2\over t}\right)^{5/4+n/2}
\label{Qdotdecay}
\end{eqnarray}
for the rate of first return to the origin. Here $_2F_1(a,b;c;z)$ denotes a hypergeometric function \cite{gr}, and the infinite series in Eq. (\ref{Qdotdecay}) converges for arbitrary $v_0^2/t$. Thus, the probability that a particle which leaves the origin with velocity $v_0$ has not yet returned after a time $t$ decays as \cite{mg}
\begin{equation}
Q(0,v_0;t)\approx{3\Gamma({1\over 4})\over 2\pi^{3/2}}\left(\displaystyle{v_0^2\over t}\right)^{1/4},\quad t\gg v_0^2\thinspace.\label{Qdecay1}
\end{equation}

\subsection{Multiple returns to the origin}\label{lachal}

The probability density $v P_n(0,-v;0,v_0;t)$ for the $n^{\rm th}$ return of a randomly accelerated particle to the origin, which is determined by the hierarchy
\begin{eqnarray}
P_n(0,-v;0,v_0;t)&=&\int_0^{t} dt_1\int_0^\infty dv_1\; P(0,-v;0,v_1;t-t_1)\nonumber\\&&\qquad\qquad \times\;
v_1P_{n-1}(0,-v_1;0,v_0;t_1),\label{la1}
\end{eqnarray}
was analyzed by Lachal \cite{al}. Here, $v$ and $v_0$ are positive, and to avoid complications with signs, we assume that the particle is instantaneously and elastically reflected back onto the positive real axis each time it arrives the origin and just before the $n^{\rm th}$ arrival has speed $v$ and velocity $-v$. Writing Eq. (\ref{la1}) in terms of Laplace transforms, making use of Eqs. (\ref{kl1})-(\ref{mck2}), and performing the inverse Laplace transformation, one obtains
\begin{eqnarray}
&&\tilde{P}_n(0,-v;0,v_0;s)\nonumber\\ &&\ \ ={2\over\pi^2v v_0}\int_0^\infty d\gamma\thinspace\thinspace{\gamma\sinh(\pi\gamma)\over\left[2\cosh(\pi\gamma/3)\right]^n}\thinspace K_{i\gamma}\left(\sqrt{4s}\thinspace v\right)K_{i\gamma}\left(\sqrt{4s}\thinspace v_0\right)\label{la2}
\end{eqnarray}
and
\begin{eqnarray}
&&P_n(0,-v;0,v_0;t)={1\over\pi^2 vv_0 t}\thinspace
\exp\left[-\left(v^2 + v_0^2\right)/t\right]\nonumber\\ &&\qquad\quad\quad\quad\quad\times\int_0^\infty d\gamma\thinspace{\gamma\sinh(\pi\gamma)\over\left[2\cosh(\pi\gamma/3)\right]^n}\thinspace K_{i\gamma}\left(2vv_0/t\right).
\label{la3}
\end{eqnarray}

The rate $I_n(v_0,t)=\int_0^\infty dv\thinspace vP_n(0,-v;x_0,v_0;t)$ of $n^{\rm th}$ return to the origin  and its Laplace transform follow from Eqs. (\ref{la2}) and (\ref{la3}) and are given by \cite{al}
 \begin{eqnarray}
 &&\tilde{I}_n\left(v_0,s\right)={1\over\pi v_0 s^{1/2}}\int_0^\infty d\gamma\thinspace\thinspace{\gamma\sinh(\pi\gamma/2)\over\left[2\cosh(\pi\gamma/3)\right]^n}\thinspace K_{i\gamma}\left(\sqrt{4s}\thinspace v_0\right),\label{la4}\\
 &&I_n\left(v_0,t\right)={e^{-v_0^2/2t}\over 2\pi^{3/2}v_0 t^{1/2}}\thinspace
\int_0^\infty d\gamma\thinspace{\gamma\sinh(\pi\gamma/2)\over\left[2\cosh(\pi\gamma/3)\right]^n}\thinspace K_{i\gamma/2}\left(v_0^2/2t\right).
\label{la5}
 \end{eqnarray}

 The first return rate $I_1(v_0,t)$ given by Eq. (\ref{la5}) has the series expansion (\ref{Qdotdecay}) and
 decays  as $t^{-5/4}$ for $t>>v_0^2$. With the approach of Ref. \cite{dsgl}, one finds that $I_2$ decays as $t^{-5/4}\ln t$, slower by a factor $\ln t$, and for larger $n$ there are higher powers of $\ln t$. The first and second return rates are compared in Fig. 1.\\
 
 \begin{figure}[H]
\begin{center}
\captionsetup{width=.8\linewidth}
\includegraphics[width=4.5in]{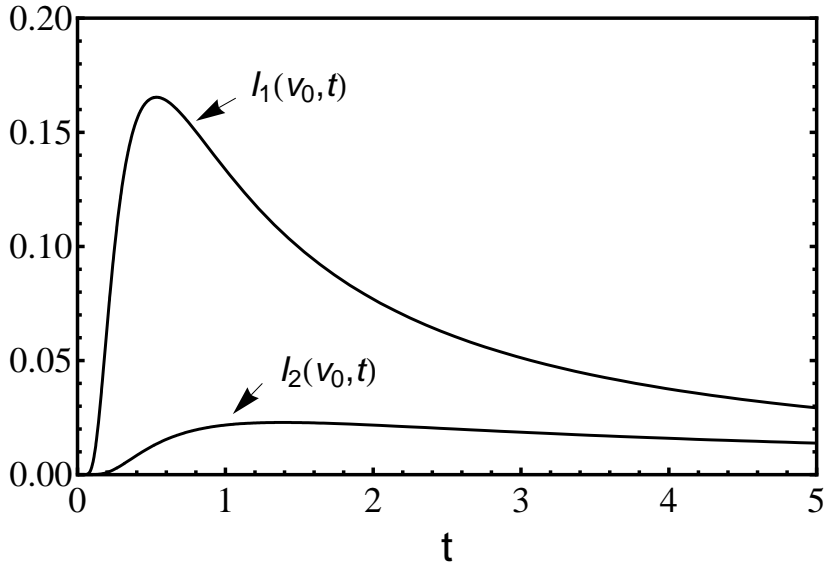}
\caption{Time dependence of the first and second return rates to the origin, as given by 
Eq. (\ref{la5}), for initial velocity $v_0=1$. The integrated return rate or area under each curve is 1. As discussed below Eq. (\ref{la5}), $I_1$ decays as $t^{-5/4}$ and $I_2$ more slowly by a factor $\ln t$. The two curves cross at $t\approx 50$, with $I_2 > I_1$ after that.}
\end{center}
\label{fig1}
\end{figure}

The time-integrated $n^{\rm th}$ arrival rate $q_n(v_0,t)=\int_0^t dt_1\thinspace I_n\left(v_0,t_1\right)$ follows from Eqs. (\ref{la4}) and (\ref{la5}) and has the form \cite{al}
 \begin{equation}
 q_n(v_0,t)={t e^{-v_0^2/2t}\over 2\pi v_0^2}\thinspace
\int_0^\infty d\gamma\thinspace{\gamma\sinh(\pi\gamma/2)\over\left[2\cosh(\pi\gamma/3)\right]^n}\thinspace W_{-1,i\gamma/2}(v_0^2/t),
\label{la6}
\end{equation}
where $W_{\lambda,\mu}(z)$ Whittaker's function \cite{as,gr}. In the limit $t\to\infty$, $q_n(v_0,t)\to 1$. 

Since $q_n(v_0,t)$ is the probability that a particle which begins at the origin with velocity $v_0$ makes $n$ or more returns to the origin in a time $t$, the probability $p_n(v_0,t)$ that it makes $n$ and only $n$ returns is given by
\begin{equation}
p_n(v_0,t)=q_n(v_0,t)-q_{n+1}(v_0,t)\label{pn(v0,t)}
\end{equation}
The distribution $p_n(v_0,t)$ has been analyzed in detail by Majumdar and Bray \cite{mb98} and Schehr and Majumdar \cite{sm07} within the framework of Gaussian stationary processes and by De Smedt et al. \cite{dsgl} on the basis of Eqs. (\ref{la3})-(\ref{pn(v0,t)}). These studies also utilize results for random acceleration on the half line with partial absorption at the boundary, considered in Sec. \ref{pr}. For $t\gg v_0^2$, $p_n(v_0,t)$ decays as $t^{-1/4}(\ln t)^n$ and $\langle n\rangle$ increases as $\ln t$.

\subsection{Solution of the Fokker-Planck equation}\label{marshallwatson}

In 1985 Marshall and Watson \cite{mw} derived the Laplace transform
$\tilde{P}(x,v;x_0,v_0;s)$ of the
propagator for the Langevin equation for Brownian motion, $\ddot{x}+\lambda
\dot{x}=\eta(t),$ with an absorbing boundary at the origin. This was accomplished by
solving the corresponding Fokker-Planck equation in terms of a special set of basis
functions, constructed explicitly in their paper, which vanish at $x=0$ for positive
$v$ but not negative $v$, so as to fulfill the boundary condition (\ref{absorbbc}).

In the limit in which the damping constant $\lambda$ approaches zero, the
Langevin equation reduces to the random acceleration process (\ref{eqmo}),
and the result of Marshall and Watson simplifies to \cite{twb93}
\begin{eqnarray}
\tilde{P}(x,v;x_0,v_0;s)&=&\tilde{P}_0(x,v;x_0,v_0;s)
-{1\over 2\pi}\int_0^\infty dF\;F^{-1/6}\int_0^\infty dG\;G^{-1/6}\nonumber\\
&\times&(F+G)^{-1}\exp\left[-Fx-Gx_0-{2\over 3}s^{3/2}\left(F^{-1}+G^{-1}\right)
\right]\nonumber\\
&\times&{\rm Ai}\left(-F^{1/3}v+F^{-2/3}s\right){\rm Ai}\left(G^{1/3}v_0+
G^{-2/3}s\right),\label{twbhalfspace1}
\end{eqnarray}
\begin{eqnarray}
&&\tilde{P}_0(x,v;x_0,v_0;s)=\int_0^\infty dF\;F^{-1/3}\nonumber\\
&&\times\Big[\theta(x-x_0) e^{-F(x-x_0)}{\rm
Ai}\left(-F^{1/3}v+F^{-2/3}s\right){\rm Ai}\left(-F^{1/3}v_0+
F^{-2/3}s\right)\nonumber\\
&&+\theta(x_0-x)e^{-F(x_0-x)}{\rm Ai}\left(F^{1/3}v+F^{-2/3}s\right) {\rm
Ai}\left(F^{1/3}v_0+F^{-2/3}s\right)\Big].\label{twbhalfspace2}
\end{eqnarray}
Here ${\rm Ai}(z)$ is the Airy function \cite{as}, and $\tilde{P}_0(x,v;x_0,v_0;s)$ is the Laplace transform of the free propagator
(\ref{P0}) for motion on the unbounded $x$ axis.

Equations (\ref{twbhalfspace1}) and (\ref{twbhalfspace2}) specify the unique superposition of the separable solutions mentioned below  Eq. (\ref{probcons}) which satisfies the
Fokker-Planck equation (\ref{fpeq}) with the initial condition (\ref{initcond}) and
the absorbing boundary condition (\ref{absorbbc}).
That $\tilde{P}(0,v;x_0,v_0;s)$ vanishes for $v>0$, in accordance with Eq. (\ref{absorbbc}), is not immediately obvious from Eqs. (\ref{twbhalfspace1}) and (\ref{twbhalfspace2}) but follows from the identity
\begin{eqnarray}
{1\over F^{1/6}}{\rm Ai}\left(F^{1/3}v+F^{-2/3}s\right)
&=&{1\over 2\pi}\int_0^\infty {dG\over F+G}\exp\left[-\textstyle{2\over 3}s^{3/2}\left(F^{-1}+G^{-1}\right)\right]\nonumber\\&&\qquad\times{1\over G^{1/6}}{\rm Ai}\left(-G^{1/3}v+
G^{-2/3}s\right).\label{identity}
\end{eqnarray}
 This relation only holds for $v>0$ and may be derived by rewriting the right-hand side of Eq. (\ref{identity}) as an integral in the complex $G$ plane with contour surrounding the positive $G$ axis, deforming the contour so it surrounds the pole at $G=-F$, and then invoking Cauchy's residue theorem.

Expressions (\ref{twbhalfspace1}) and (\ref{twbhalfspace2}) provide the extension of McKean's results (\ref{mck2}) and (\ref{mck3}) for the half space propagator with $x=x_0=0$ to arbitrary points $x$ and $x_0$ on the positive $x$ axis. Inverting the Laplace transform in Eq. (\ref{twbhalfspace1}) analytically appears out of the question, but for short times $P(x,v;x_0,v_0;t)$ approaches the free propagator (\ref{P0}), and for long times $t\gg x^{2/3}$, $x_0^{2/3}$, $v^2$, and $v_0^2$,

\begin{eqnarray}
P(x,v;x_0,v_0;t)&\approx&3^{5/3}\pi^{-3/2}t^{-5/2}\left(xx_0\right)^{1/6}\nonumber\\ &&\quad\times \;U\left(-{1\over 6},{2\over 3},-{v^3\over 9x}\right)U\left(-{1\over 6},{2\over 3},{v_0^3\over 9x_0}\right),\label{Plarget}
\end{eqnarray}
as shown in Ref. [{\cite{twb93}, where $U(a,b,c)$ is Kummer's confluent hypergeometric function \cite{as}.

The Laplace transform of the survival probability follows from Eqs. (\ref{Qdef}), (\ref{twbhalfspace1}), and (\ref{twbhalfspace2}) and is given by \cite{twb93}
\begin{eqnarray}
\tilde{Q}(x_0,v_0;s)&=&s^{-1}-\int_0^\infty dF\thinspace F^{-5/3}e^{-Fx_0}{\rm Ai}\left(F^{1/3}v+F^{-2/3}s\right)\nonumber\\
&\times&\left[1+{1\over 4\pi^{1/2}}\Gamma\left(-\textstyle{1\over 2},{2\over 3}F^{-1}s^{3/2}\right)\right].\label{Qtilde}
\end{eqnarray}
The small $s$ behavior of $\tilde{Q}(x_0,v_0;s)$, which diverges as $s^{-3/4}$, implies the long time behavior
\begin{equation}
Q(x_0,v_0;t)\approx {3^{4/3}\Gamma({1\over 4})\over 2\pi^{3/2}}\left({x_0^{2/3}\over t}\right)^{1/4}U\left(-{1\over 6},{2\over 3},{v_0^3\over 9x_0}\right)\label{Qdecay2}
\end{equation}
for $t\gg x_0^{2/3}$, $t\gg v_0^{2}$.
Expression (\ref{Qdecay2}) extends the McKean-Goldman result (\ref{Qdecay1}) for $x_0=0$ to arbitrary $x_0$.

Due to the slow $t^{-1/4}$ decay in Eq. (\ref{Qdecay2}), the average time of first arrival at the origin,
\begin{equation}
T(x_0,v_0)=\int_0^\infty dt\thinspace t\left[-{\partial\over\partial t}\thinspace Q(x_0,v_0;t)\right]=\int_0^\infty dt\thinspace Q(x_0,v_0;t),\label{exittime}
\end{equation}
is infinite.

The $t^{-1/4}$ dependence of the survival probability in Eqs. (\ref{Qdecay1}) and (\ref{Qdecay2}) also follows from the argument, see, e.g., \cite{bms,sm}, that $Q$ decays as $N^{-1/2}$ for large $N$, where $N$ is the number of zero crossings of the velocity, in accordance with the Sparre-Anderson theorem \cite{sa}, and that $N$ scales as $t^{1/2}$. Sinai \cite{ys} has given a rigorous derivation of the $t^{-1/4}$ behavior and other first-passage properties along these lines. For a derivation \cite{twb00} that leads to the asymptotic form (\ref{Qdecay2}) in just a few steps (but does not determine the multiplicative constant) see Sec. \ref{pr}.

\subsection{Particle subject to a constant force and a random force}\label{gravity}
For a particle subject to a constant force, such as gravity, in addition to a random force, and governed by the equation of motion $\ddot{x}=g+\eta(t)$, the Fokker-Planck equation (\ref{fpeq}) is replaced by
\begin{equation} \left({\partial\over\partial t}+v{\partial\over\partial
x}+g{\partial\over
\partial v}-{\partial^2\over
\partial v^2}\right)P_g(x,v;x_0,v_0;t)=0\thinspace,\label{fpg}.
\end{equation}
Here, by rescaling of $x$ and $t$, we have chosen $g=\pm 1$ and $2\Lambda=1$
with no loss of generality. The initial condition for $P_g(x,v;x_0,v_0;t)$ is the same as in Eq. (\ref{initcond}), and for studying the first exit from the positive real axis, the same absorbing boundary condition (\ref{absorbbc}) applies.
Keeping this in mind and comparing Eqs. (\ref{fpeq}) and (\ref{fpg}), we see that the propagators with and without the constant force $g$ are related by
\begin{equation}
P_g(x,v;x_0,v_0;t)=\exp\left[\textstyle{1\over
2}g(v-v_0)-\textstyle{1\over 4}t\right]P(x,v;x_0,v_0;t).\label{Pgamma}
\end{equation}

Making use of Eq. (\ref{Pgamma}) and expressions (\ref{mck3}) and (\ref{twbhalfspace1}) for the propagator on the right-hand side, Burkhardt \cite{twb08} has studied the first exit from the positive $x$ axis of a particle subject to a constant force and a random force. In contrast to the long time behavior discussed in connection with Eqs. (\ref{Qdecay2}) and ({\ref{exittime}), the survival probability $Q_g(x_0,v_0;t)$ for $g=1$, corresponding to a constant force driving the particle in the $+x$ direction away from the origin, does not vanish for $t\to\infty$, and in the case $g=-1$ of a constant force toward the origin, the mean time $T_g(x_0,v_0)$ of arrival at the origin is finite.

\subsection{Partially absorbing, inelastic boundary}\label{pr}
Consider a randomly accelerated particle on the half line $x>0$, which, on reaching the origin, is absorbed with probability $1-p$ and reflected inelastically with probability $p$ and velocity $-rv$, where $r$ is the coefficient of restitution. The probability $Q(x_0,v_0;t)$ that the particle has not yet been absorbed after a time $t$ satisfies the backward Fokker-Planck equation (\ref{fpeqQ}) with initial condition (\ref{initcondQ}) and boundary condition
\begin{equation}
Q(0,-v_0;t)=pQ(0,rv_0;t),\quad v_0>0.\label{Qbc}
\end{equation}
Burkhardt \cite{twb00} showed that $Q(x_0,v_0;t)$ decays asymptotically as $t^{-\phi}$ for large $t$, with persistence exponent $\phi(p)$ given by
\begin{equation}
2\sin\left[{\pi\over 6}(1-4\phi)\right]=pr^{2\phi}.\label{phi(pr)}
\end{equation}

Making use of Eqs. (\ref{la3})-(\ref{pn(v0,t)}), De Smedt et al. \cite{dsgl} confirmed the persistence exponent (\ref{phi(pr)}) and derived the functional form of the survival probability $\tilde{Q}(0,v_0;s)$ for the case $r=1$ of absorption with probability $1-p$ and elastic reflection with probability $p$. This quantity has the expansion
\begin{equation}
Q(0,v_0;t)=\sum_{n=0}^\infty p^np_n(v_0,t)\label{genfunc}
\end{equation}
in terms of the probability $p_n(v_0,t)$ of $n$ returns in a time $t$ introduced in Eq. (\ref{pn(v0,t)}) and can be used as a generating function in calculating the moments of $n$. The studies \cite{mb98,sm07,dsgl} of the distribution $p_n(v_0,t)$ cited at the end of Sec. \ref{lachal} are carried out in the context of random acceleration with partial absorption and make use of Eqs. (\ref{phi(pr)}) and (\ref{genfunc}).

A short route to the persistence exponent (\ref{phi(pr)}), following Ref. \cite{twb00}, is as follows: Scaling invariance restricts $Q(x_0,v_0;t)$ to a function of two scale independent variables $v_0^3/x_0$ and $t^3/x_0^2$. This and the assumption of a $t^{-\phi}$
decay imply
\begin{equation}
 Q(x_0,v_0;t)\approx \left(tx_0^{-2/3}\right)^{-\phi}{\cal F}\left(v_0^3/9x_0\right).
\end{equation}
Substituting this in the Fokker-Planck equation
(\ref{fpeqQ}) and dropping all terms which decay faster than $t^{-\phi}$ leads to the confluent hypergeometric \cite{as} differential equation  $\left[z\partial_z^2 +
\left({2\over 3}-z\right)\partial_z+{2\over 3 }\phi z\right]{\cal F}(z)=0$. The solution which is finite for $v_0\to\infty$ is given by ${\cal
F}=C U\left(-{2\over 3}\phi,{2\over 3},v_0^3/9x_0\right)$, where $C$ is a
constant and $U(a,b,z)$ is Kummer's function \cite{as}.
Imposing the boundary condition (\ref{Qbc}) leads directly to expression (\ref{phi(pr)})
for the persistence exponent $\phi$. For $p=0$ the first-passage  probability given in Eq. (\ref{Qdecay2}) is reproduced, apart from the multiplicative constant.

A randomly accelerated particle moving between two walls from which it is reflected inelastically is expected to reach a steady-state, in which the kinetic energy gain due to the random force is lost in the inelastic collisions. The question whether there is ``inelastic collapse" or localization of the particle at the boundaries has been considered by several authors, see \cite{csb}-\cite{kb} and references therein. According to Ref. \cite{bk1}, the equilibrium rate of collision with the boundaries is infinite for coefficient of restitution $r<r_c=\exp\left(-\pi/\sqrt 3\right)=0.163$..$\thinspace$, as first pointed out by Cornell et al. \cite{csb}, but this does not lead to localization of the particle at the boundary.

Similar conclusions were reached \cite{bk2} for a particle which is accelerated on the half line $x>0$ by both a random force and a constant force directed to the origin, where it is reflected inelastically. The collision rate at the origin is infinite for $r<r_c=\exp\left(-\pi/\sqrt 3\right)$, but this does not lead to localization at the origin. For two values $r={1\over 2}$ and ${1\over 3}$, the steady state distribution function $P_{\rm eq}(x,v)$ was calculated explicitly.

\subsection{Windy cliff problem}\label{windycliff}
Consider a particle in the half plane $(x,y)$ with $x>0$ which moves diffusively in the $y$ direction and and is transported in the $x$ direction by a flow field $f(y)$ according to
\begin{equation}
\dot{y}=\eta(t),\qquad \dot{x}=f(y).\label{windycliff}
\end{equation}
In the case $f(y)=y$ of uniform shear flow, Eq. (\ref{windycliff}) is equivalent to the random acceleration process $\ddot{x}=\eta(t)$, and the probability that the particle has not yet reached the boundary $x=0$ after a time $t$ decays asymptotically as $t^{-1/4}$ for long times, as in Eq. (\ref{Qdecay2}).

Redner and Krapivsky \cite{rk} have shown that for the flow field $f(y)=v_0\thinspace{\rm sgn}(y)$, the survival probability also decays as $t^{-1/4}$ , and Bray and Gonos \cite{bg} have extended this result to any $f(y)$ which is an odd function of $y$ and has the same sign as $y$. Results for an even larger class of flow fields and for partially absorbing boundaries are given in \cite{bm,bms}.

\section{First Exit from the Interval $0<x<1$}\label{finiteinterval}
\subsection{Mean Exit Time}\label{escapetime}
Unlike the mean exit time (\ref{exittime}) from the positive $x$ axis, the mean exit time $T(x_0,v_0)$ of a randomly accelerated particle from a line segment of finite length is finite. Rescaling of both $x$ and $t$ allows us to set the segment length equal to 1 and $2\Lambda=1$ in Eqs. (\ref{eqmo}) and (\ref{fpeq}), and with no loss of generality we consider the interval $0<x<1$.
From Eq. (\ref{exittime}) and the backward Fokker-Planck equation (\ref{fpeqQ})
for the survival probability, we obtain the inhomogeneous partial differential equation
\begin{equation}
\left(v_0{\partial\over\partial x_0}+{\partial^2\over\partial v_0^2}
\right)T(x_0,v_0) =-1,\label{Tfpeq}
\end{equation}
for the mean exit time, to be solved with the boundary conditions
\begin{eqnarray}
T(0,v_0)&=&0,\quad v_0<0,\label{Tbc1}\\
T(1,v_0)&=&0,\quad v_0>0.\label{Tbc2}
\end{eqnarray}
This differential equation looks simpler than the Fokker-Planck equation
(\ref{fpeqQ}), since there is no time derivative, but a complicating feature is that
there are two boundaries at which the particle can exit. The system of equations (\ref{Tfpeq})-(\ref{Tbc2}) was solved by Franklin and Rodemich \cite{fr} and Masoliver and Porr\`a \cite{mp}, and the mean escape time is
\begin{eqnarray}
T(x_0,v_0)&=&T(1-x_0,-v_0)={1\over 3}(\pi v_0)^{1/2}\int_0^{1-x_0} dy\thinspace
{e^{-v_0^3/18y}\over y^{1/2}}I_{-1/6}\left({v_0^3\over 18 y}\right)
\nonumber\\&&+{2^{-7/3}3^{5/6}\over\Gamma({2\over 3})^2}
\int_{x_0}^1 dy\thinspace {e^{-v_0^3/9(y-x_0)}\over
(y-x_0)^{2/3}}[y(1-y)]^{-1/6}\nonumber\\&&\times\left[_2F_1(1,-\textstyle{2\over 3};{5\over
6};1-y)-\thinspace_2F_1(1,-{2\over 3};{5\over 6};y)\right]\label{mp},\quad v_0>0.
\end{eqnarray}

\subsection{Splitting Probabilities}\label{splitprob}
The probabilities $q_0(x_0,v_0)$ and $q_1(x_0,v_0)$ that the particle makes its first exit from the line segment $0<x<1$ at $x=0$ and $x=1$, respectively, are determined by the sum rule $q_0(x_0,v_0)+q_1(x_0,v_0)$, the reflection symmetric property $q_0(x_0,v_0)=q_1(1-x_0,-v_0)$, and the differential equation
\begin{equation}
\left(v_0{\partial\over\partial x_0}+{\partial^2\over\partial v_0^2}
\right)q_0(x_0,v_0)=0,\label{qfpeq}
\end{equation}
with boundary condition
\begin{equation}
q_0(0,v_0)=1, \quad v_0<0.
\end{equation}
The solution, obtained by Bicout and Burkhardt \cite{bb1}, is
\begin{eqnarray}
q_0(x_0,v_0)&=&1-q_0(1-x_0,-v_0)\nonumber\\&=&{1\over 2\pi}\int_{x_0}^1 dy\thinspace
{e^{-v_0^3/9(y-x_0)}\over (y-x_0)^{2/3}}\left[y(1-y)\right]^{-1/6},\quad v_0>0.
\end{eqnarray}

\subsection{Long Time Behavior}\label{longescapetime}
In the long time limit, the probability density for propagation of the particle from $x_0,v_0$ to $x,v$ without leaving the line segment $0<x<1$ decays as
\begin{equation}
P(x,v;x_0,v_0;t)\approx\psi_0(x,v)\psi_0(x_0,-v_0)e^{-E_0t},\label{Plongtime}
\end{equation} where $E_0$ is
the eigenvalue with the smallest real part of the time-independent
Fokker-Planck equation
\begin{equation}
\left(-E+v{\partial\over\partial x}-{\partial^2\over\partial v^2}\right)\psi(x,v) =0\thinspace ,\label{tindepfpeq}
\end{equation}
with the absorbing boundary condition $\psi(0,v)=\psi(1,-v)=0$ for $v>0$, analogous to Eq. (\ref{absorbbc}). From Eqs. (\ref{Qdef}) and (\ref{Plongtime}) it is clear that the survival probability $Q(x_0,v_0;t)$ also decays as $e^{-E_0 t}$. Solving numerically an integral equation for $E_0$ and $\psi_0$ that arises in an
exact analytic approach, Burkhardt \cite{twb97} obtained $E_0=1.3904$ to 5 significant figures, a result which is in excellent agreement with precision simulations. As discussed below in Sec. \ref{channel}, this number is of interest in connection with the equilibrium statistical mechanics of semi-flexible polymers confined in narrow channels.

\section{Extreme Statistics of the Random Acceleration Process}\label{extreme}

Consider the family of possible trajectories of a randomly accelerated particle
which lead from $x_0,v_0$ to $x_1,v_1$ in a time $t=T$. Each trajectory $x(t)$ reaches a maximum value of $x$ at some time in the interval $0\leq t\leq T$. The probability that the maximum value lies between $X$ and $X+dX$ may be expressed as ${\cal P}(X,T|x_1,v_1;x_0,v_0)\thinspace dX$. Probability distributions such as ${\cal P}$ play a central role in the field of extreme value statistics \cite{ejg,jg,sc,chbs}. We will see that the extreme value  and first-passage statistics of random acceleration are closely related.

We begin with the relation
\begin{equation}
{\cal P}={\partial{\cal F}\over\partial X}\;,\label{calFtoP}
\end{equation}
between the desired distribution and the cumulative probability
${\cal F}(X,T|x_1,v_1;x_0,v_0)$ that the maximum
displacement of a trajectory from $x_0,v_0$ to $x_1,v_1$ in time $T$ is less than
$X$. From invariance under the translation $x\to x-X$ and the reflection $x\to -x$, $v\to -v$, ${\cal F}(X,T|x_1,v_1;x_0,v_0)$ is the same as the probability of propagation from $X-x_0,-v_0$ to $X-x_1,-v_1$, without ever leaving the positive $x$ axis. Thus, in the regime $X>{\rm max}\left(x_1,x_2\right)$ in which ${\cal F}$ does not vanish identically,
\begin{eqnarray}
{\cal F}(X,T|x_1,v_1;x_0,v_0) &=&{P(X-x_1,-v_1;X-x_0,-v_0;T)\over P_0(X-x_1,-v_1;X-x_0,-v_0;T)}.\label{calF1}
\end{eqnarray}
Here $P(x,v;x_0,v_0;t)$ is the half-line propagator with absorbing boundary condition at the origin, discussed at length in Sec. \ref{halfline}, and $P_0(x,v;x_0,v_0;t)$ is the corresponding propagator in the absence of boundaries, given explicitly in Eq. (\ref{P0}).

The asymptotic form of ${\cal F}(X,T|x_1,v_1;x_0,v_0)$ for large $T$ follows directly from Eqs. (\ref{P0}), (\ref{Plarget}), and (\ref{calF1}), and for general $T$, ${\cal F}(X,T|x_1,v_1;x_0,v_0)$
can be evaluated by numerical inversion of the Laplace transform (\ref{twbhalfspace1}).

It is simple to derive analogous extreme distributions for other boundary conditions of
interest. For example, the probability that the maximum displacement
of a trajectory which begins at $x_0,v_0$ does not exceed $X$ in a time $T$, with no
restrictions on $x_1,v_1$, is given by
\begin{eqnarray}
{\cal F}(X,T|x_0,v_0)&=&{\int_{-\infty}^\infty dv_1\int_{-\infty}^X dx_1\;
P(X-x_1,-v_1;X-x_0,-v_0;T)\over \int_{-\infty}^\infty dv_1 \int_{-\infty}^\infty
dx_1\;P_0(X-x_1,-v_1;X-x_0,-v_0;T)}\nonumber\\
&=& Q(X-x_0,-v_0;t),\label{calF3}
\end{eqnarray}
where we have used the normalization condition (\ref{norm}), and where $Q$ is the survival probability introduced in Eq. (\ref{Qdef}), with Laplace transform and asymptotic form for large $t$ given in Eqs. (\ref{Qtilde}) and (\ref{Qdecay2}), respectively. Although the distribution ${\cal P}(X,T|x_0,v_0)$ defined by Eqs. (\ref{Qtilde}), (\ref{calFtoP}), and (\ref{calF3}) has not been evaluated analytically, Reymbaut et al. \cite{rmr} have derived all its moments $\langle(X-x_0)^n\rangle_{_T}$ for $v_0=0$. The first two are given by
\begin{equation}
\langle X-x_0\rangle_{_T}=\sqrt{{3\over 8\pi}}\thinspace T^{3/2},\quad
\langle(X-x_0)^2\rangle_{_T}=\left({2\over 3}-{103\over 384}\sqrt{{3\over 2}}\right)T^3.
\end{equation}

The extreme displacement of trajectories which start at the
origin, return after a time $T$, and are periodic, with $x(t)=x(t+T)$ and $x(0)=0$ is analyzed in Ref. \cite{twbbudapest}.
In analogy with Eqs. (\ref{calF1}} and (\ref{calF3}), the probability
such a periodic trajectory has maximum displacement less than $X$ is given by
\begin{equation}
{\cal F}_{\rm per}(X,T)={\int_{-\infty}^\infty dv\;P(X,v;X,v;T)\over
\int_{-\infty}^\infty dv\;P_0(X,v;X,v;T)}.\label{calF3per}
\end{equation}
Evaluating the right hand side of Eq. (\ref{calF3per}) using Eqs. (\ref{P0}) and
(\ref{twbhalfspace1}) and differentiating with respect to $X$, in accordance with
Eq. (\ref{calFtoP}), leads to the extreme value distribution \cite{twbbudapest}
\begin{equation}
{\cal P}_{\rm per}(X,T)={2^{4/3}3^{5/6}\over \sqrt{\pi}X}\left({X\over
T^{3/2}}\right)^{2/3}e^{-48X^2/T^3}\;U(-{1\over   6},{2\over 3},{48X^2\over
T^3}).\label{extremeperiodic}
\end{equation}
Here again, $U(a,b,z)$ is Kummer's confluent hypergeometric function \cite{as}. For more
details and a discussion of the extreme statistics of the process $d^n
x/dt^n=\eta(t)$ for general integer and non-integer $n$, not just the case $n=2$ of random acceleration, see \cite{twbbudapest}.

The distribution ${\cal P}_g(X,v_0)$ of the maximum height $X$ attained by a particle which is thrown vertically upwards with velocity $v_0$ and is subject to both a random force and a downward gravitational force $g=-1$ is studied in Ref. \cite{twb08}. As in the examples considered above, the extreme distribution follows from the ratio of the propagators, discussed in Sec. \ref{gravity}, for motion on the half line and entire line and is given by \cite{twb08}
\begin{eqnarray}
&&{\cal P}_g(X,v_0)=\thinspace{e^{v_0/2}\over\sqrt{2\pi}}\int_0^\infty dF\thinspace
F^{-1/6}\nonumber\\
&&\qquad\times\thinspace\exp\left(-\thinspace{1\over 12F}-FX\right){\rm
Ai}\left(-F^{1/3}v_0+\textstyle{1\over 4}\displaystyle\thinspace
F^{-2/3}\right)\label{calPg}
\end{eqnarray}
for positive $v_0$. In the limit of large $v_0$, the mean height and standard deviation vary according to   $\langle X\rangle\approx{1\over 2}v_0^2+v_0$ and as $\sigma\approx\left({2\over 3}v_0^3\right)^{1/2}$, as compared to the familiar maximum height ${1\over 2}v_0^2$ in the absence of the random force.

Finally we note that Reymbaut et al. \cite{rmr} make use of extreme value statistics in their analysis of the convex hulls of trajectories of randomly accelerated particles in two dimensions, which begin at the origin with velocity zero and continue for a time $T$. For the mean values of the perimeter $L$ and area $A$ of the hull, they obtain

 \begin{equation}
\langle L\rangle_{_T}=\sqrt{{3\pi\over 2}}\thinspace T^{3/2},\quad
\langle A\rangle_{_T}={5\pi\over 192}\sqrt{{3\over 2}}\thinspace T^3.
\end{equation}

\begin{figure}[H]
\begin{center}
\captionsetup{width=.8\linewidth}
\includegraphics[width=5.5in,clip=true,trim= 0 450 0 100 mm]{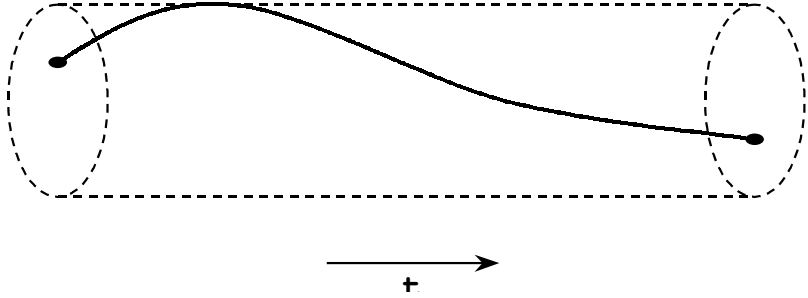}
\caption{The curve may be interpreted as the trajectory $\vec{r}(t)$ of a randomly accelerated particle moving in two dimensions, plotted as a
function of $t$, or as the configuration of a tightly confined semi-flexible polymer in
a channel.}
\end{center}
\label{fig2}
\end{figure}

\section{Applications}\label{applications}

\subsection{Semi-flexible polymer in a cylindrical channel}\label{channel}
Consider a long, semi-flexible polymer or worm-like chain fluctuating in equilibrium in a narrow cylindrical channel. If the width of the channel is much smaller than the persistence length $P$, typical polymer configurations hardly deviate from straight lines and correspond, as
shown in Fig. 2, to single valued functions $\vec{r}(t)$, where $(x,y,t)$ are
Cartesian coordinates, and $\vec{r}=(x,y)$ specifies the transverse displacement of
the polymer from the symmetry axis or $t$ axis of the tube. The
bending energy ${\cal H}={1\over 2}\kappa\int ds\left(d\hat{\tau}/ds\right)^2$, where $s$ is the arc length and $\hat{\tau}$ is a unit tangent vector, simplifies to to ${\cal H}={1\over 2}\kappa\int dt\left(d^2\vec{r}/dt^2\right)^2$, and the polymer partition function is given by
\begin{equation}
Z(\vec{r},\vec{v};\vec{r}_0,\vec{v}_0;t)=\int D^2r\exp\left[-{P\over
2}\int_0^t\left({d^2\vec{r}\over dt^2}\right)^2\right],\label{polpartfunc}
\end{equation}
where $P=\kappa/k_BT$, and satisfies the Fokker-Planck equation
\begin{equation}
\left({\partial\over\partial t}+\vec{v}\cdot{\bf\nabla_r}-{1\over 2P} \nabla_{\bf
v}^2\right)Z(\vec{r},\vec{v};\vec{r}_0,\vec{v}_0;t) =0 \label{fpeqpol}.
\end{equation}

Since discontinuities in slope of the polymer at the boundary cost an infinite bending energy and are suppressed, $Z(\vec{r},\vec{v};\vec{r}_0,\vec{v}_0;t)$ vanishes as $\vec{r}$ approaches the wall of the tube for $\hat{n}\cdot\vec{v}>0$, where $\hat{n}$ is normal to the wall and directed toward the interior of the tube.

From comparing the path integrals (\ref{pathintegral}) and (\ref{polpartfunc}), the Fokker-Planck equations (\ref{fpeq}) and (\ref{fpeqpol}), and the boundary condition (\ref{absorbbc}) with the one in the preceding paragraph, one concludes \cite{twb97} that the polymer partition function is identical with the probability that a randomly accelerated particle moving in two dimensions propagates from $(\vec{r}_0,\vec{v}_0)$ to $(\vec{r},\vec{v})$ without leaving a two-dimensional domain corresponding to the cross section of the channel. As illustrated in Fig. 2, each of the possible trajectories of the randomly accelerated particle may be interpreted as a polymer configuration.

In the long-polymer limit, the partition function (\ref{polpartfunc}) has the asymptotic form
$Z\approx \psi_0\left(\vec{r},\vec{v}\right)\psi_0\left(\vec{r_0},-\vec{v_0}\right)e^{-E_0 t}$,
where $E_0$ is the smallest eigenvalue of the $t$-independent Fokker-Planck equation and $\psi_0\left(\vec{r},\vec{v}\right)$ is the corresponding eigenfunction. Thus, we interpret $f= k_BT E_0$ as the free energy per unit length of confinement of the polymer. According to the correspondence in Fig. 2, the probability that the randomly accelerated particle has not yet left the two-dimensional domain corresponding to the channel cross section in a time $t$ decays as $e^{-E_0 t}$.

For a tightly confined polymer polymer in a channel with a rectangular cross section with edges $D_1$ and $D_2$, the partition function (\ref{polpartfunc}) factors in the form $Z(x,v_x;x_0,v_{0x};t)Z(x,v_x;x_0,v_{0x};t)$, and the value of $E_0$ follows from the numerical result for random acceleration in one dimension quoted just below Eq. (\ref{tindepfpeq}). Going from dimensionless variables back to the original variables, one obtains \cite{twb97}
\begin{equation}
f\approx A_\Box{kT\over P^{1/3}}\left({1\over
D_1^{2/3}}+
{1\over D_2^{2/3}}\right)\label{frectangle}
\end{equation}
for the free energy of confinement per unit length, with $A_\Box=2^{-1/3}(1.3904)=1.1036.$

By performing computer simulations of randomly accelerated motion, one can study properties of tightly confined polymers in channels, using the correspondence illustrated in Fig. 2. The estimates $A_\Box=1.1038\pm 0.0006$ for the amplitude in Eq. (\ref{frectangle}) and the corresponding amplitude $A_\circ=2.3565\pm 0.0004$ in the free energy of confinement
\begin{equation}
f\approx A_\circ{kT\over P^{1/3}D^{2/3}},\label{fcircle}\\
\end{equation}
 for a channel with a circular cross section of diameter $D$ were obtained this way \cite{bb2,ybg,byg}, and the mean value and fluctuations of the length of channel occupied by the polymer were also studied.

The dependence on the persistence length and channel dimensions in Eqs. (\ref{frectangle}) and (\ref{fcircle}), which follows from the polymer partition function (\ref{polpartfunc}), is completely consistent with scaling predictions of Odijk \cite{to}. Finally, we call attention to recent experimental studies of single biopolymers, for example DNA \cite{retal} and actin filaments \cite{kp1}-\cite{kp3}, in channels with widths smaller than or comparable with the persistence length.

\subsection{Spatial persistence in interface growth}

Majumdar and Bray \cite{bms},\cite{mb} have studied the persistence of fluctuating interfaces which evolve according to the Langevin equation
\begin {equation}
{\partial h\over\partial t}=-\left(-\nabla^2\right)^{z/2}h+\xi,\label{langevineq}
\end{equation}
where $h\left(\vec{r},t\right)$ denotes the height of the interface above point $\vec{r}$ in a $d$-dimensional hyperplane, $z$ is a dynamical exponent, and $\xi$ is Gaussian white noise with zero mean and $\langle\xi\left(\vec{r},t\right)\xi\left(\vec{r}',t'\right)\rangle=
\Lambda\delta\left(\vec{r}-\vec{r}'\right)\delta(t-t')$. The values $z=2$ and $z=4$ correspond to the well-known cases of Edward-Wilkinson and Mullins-Herring growth, respectively.

Majumdar and Bray analyzed the correlations of the quantity $g_n(x_1,x_2,x_3, ...)=\partial^n h/\partial x_1^n$ with respect to an arbitrary direction $x_1$ in the hyperplane. In the steady-state limit $t\to\infty$, the equal time pair correlation function $\langle g_n(x_1,x_2,x_3, ...)g_n(x_1',x_2,x_3, ...)\rangle$ turns out to be proportional to $\delta(x_1-x_1')$, corresponding to Gaussian white noise, for $n={1\over 2}(z-d+1)$. In particular, in the case $z=4$, $d=1$, corresponding to Mullins-Herring or Golubovi\'c-Bruinsma-Das Sarma-Tamborenea \cite{gbr},\cite{dst} growth, this implies $d^2h/dx^2=\eta(x)$, where $\eta(x)$ is Gaussian white noise with zero mean. Since this is the same as the random acceleration process (\ref{eqmo}) with variables $(h,x)$ in place of $(x,t)$, the spatial persistence of the interface and the temporal persistence of the randomly accelerated particle are the same. Thus, the probability, in steady state, that an interface $h(x)$ with height $h_0$ and slope $h_0'$ at $x_0$ does not return to the same height $h_0$ within a distance $|x-x_0|$ of $x_0$ decays as $(h_0'^2/|x-x_0|)^{1/4}$ for large distances, as in Eq. (\ref{Qdecay1}).

For more details and an analysis of other growth models and other types of persistence in interface growth, see Refs. \cite{mb,bms}.

\subsection{Other applications}

In closing we reference several other applications where first-passage properties of random acceleration play a role:

Several authors \cite{mhl}-\cite{cd} have considered the desorption or depinning  transition in a rather artificial $1+1$ dimensional model of a semi-flexible polymer fluctuating without overhangs in the half plane $x>0$, $-\infty<t<\infty$ with partition function
\begin{equation}
Z(x,v;x_0,v_0;t)=\int Dx\exp\left\{-\int_0^t dt\left[{P\over 2} \left({d^2x\over
dt^2}\right)^2+V(x)\right]\right\},\label{partfunc1+1}
\end{equation}
reminiscent of the propagator (\ref{pathintegral}) for random acceleration and the partition function (\ref{polpartfunc}) for a nearly straight polymer. Here, $x$ and $t$ are Cartesian coordinates, and $V(x)$ is a short range interaction favoring absorption of the polymer along the boundary $x=0$. The desorption transition, as the temperature is raised, appears to be first-order, in contrast to the second-order desorption transition, in both two and three spatial dimensions and in the limit of infinite contour length, of a flexible self-avoiding, semi-flexible polymer, fluctuating with overhangs included \cite{ee}.

Schwarz and Maimon \cite{sm} make use of first-passage properties of random acceleration in analyzing a model for crack propagation in elastic media.

The propagator for random acceleration on the half line with absorbing boundary condition appears in Valageas' analysis \cite{pv} of the statistical properties of the Burgers equation with Brownian initial velocity.

Convex hulls have applications \cite {rmr} in crystallography, computer image processing, ecology, and polymer statistics. Hilhorst et al. \cite{hcs} have found a connection between the random acceleration process and Sylvester's problem of convex polygons. Reymbaut et al. \cite{rmr} map the convex hull of a two-dimensional random acceleration process onto a one-dimensional problem and make use of first-passage properties and extreme statistics of random acceleration in their analysis.

Finally we note that Majumdar et al. \cite{mrz} have analyzed the distribution $p(t_m\vert T)$ of the time $t_m$ at which a particle reaches its maximum displacement if it begins at the origin with velocity zero and is randomly accelerated for a time $T$. They obtain simple analytic expressions both for integrals of Brownian bridges (trajectories constrained to return to $v=0$ in a time $T$) and integrals of free Brownian motion (no restrictions on the trajectory at $T$). For ordinary unrestricted Brownian motion beginning at the origin, both distributions $p(t_m\vert T)$ and $p(t_{\rm occup}\vert T)$, where $t_{\rm occup}$ is the total time spent on the positive $x$ axis, are given by \cite{levy,wf} L\'evy's arcsine law $p(t\vert T)=\pi^{-1}\left[t(T-t)\right]^{-1/2}$. For random acceleration, on the other hand, Majumdar et al. conclude, on the basis of computer simulations, that $p(t_{\rm occup}\vert T)$ does not coincide with either of their results for $p(t_m\vert T)$. The derivation of the exact analytical form of $p(t_{\rm occup}\vert T)$ for random acceleration is a challenging, currently unsolved problem.

\section*{Acknowledgments} It is a pleasure to thank Dominique Bicout, Alan Bray, Dieter Forster, Claude Godr\`eche, Jerrold Franklin, Richard Gawronski, Gerhard Gompper, Stanislav Kotsev, Satya Majumdar, Zoltan R\'acz, and Yingzi Yang for stimulating discussions and correspondence. I also thank Satya for useful comments on a draft of this chapter and help with references.


\begin{thebibliography}{99}
\bibitem{hr} Risken, H. (1989). {\it The Fokker-Planck Equation:
Methods of Solution and Applications}, 2nd edn., Springer.
\bibitem{twb07} Burkhardt, T. W. (2007). {\it The random acceleration process in bounded geometries}, J. Stat. Mech. P07004 (16 pp.).
\bibitem{sr} Redner, S. (2001). {\it A Guide to First-Passage Properties}, Cambridge.
\bibitem{bms} Bray, A. J., Majumdar, S. N. and Schehr, G. (2013). {\it Persistence and first-passage properties in nonequilibrium systems}, Advances in Physics {\bf 62}, 225-361.
\bibitem{mck} McKean, H. P. (1963). {\it A winding problem for a resonator driven by a white noise}, J. Math. Kyoto Univ. {\bf 2}, pp. 227-235.
\bibitem{mw} Marshall, T. W. and Watson, E. J. (1985). {\it A drop of ink falls from my pen...It comes to earth, I know not when}, J. Phys. A {\bf 18}, pp. 3531-3559.
\bibitem{as} Abramowitz, M. and Stegun, I. A. eds. (1965). {\it Handbook of Mathematical Functions}, Dover.
\bibitem{mg} Goldman, M. (1971). {\it On the first passage of the integrated Wiener process}. Ann. Math. Stat. {\bf 42}, pp. 2150-2155.
\bibitem{gr} Gradshteyn, I. S. and Ryzhik, I. M. (1980). {\it Table of Integrals, Series, and Products}, Academic.
\bibitem{kl} Erd\'elyi. A. (1953). {Tables of Integral Transforms}, Vols. 1 and 2, Bateman Manuscript Project, McGraw-Hill.
\bibitem{al} Lachal, A. (1997). {\it Les temps de passage successifs de l'int\'egrale
du mouvement brownien}, Ann. Inst. Henri Poincar\'e B {\bf 33}, pp. 1-36.
\bibitem{dsgl} De Smedt, G., Godr\`eche, C. and Luck, J. M. (2001) {\it Partial survival and inelastic collapse for a randomly accelerated particle}, Europhys. Lett. {\bf 53}, pp. 438-443.
\bibitem{mb98} Majumdar, S. N. and Bray A. J. (1998). {\it Persistence with partial survival}, Phys. Rev. Lett. {\bf 81}, pp. 2626-2629.
\bibitem{sm07} Schehr, G. and Majumdar, S. N. (2007). {\it Statistics of the number of zero crossings: From random polynomials to the diffusion equation}, Phys. Rev. Lett. {\bf 99}, 060603 (4 pp.).
\bibitem{twb93} Burkhardt, T. W. (1993). {\it Semiflexible polymer in the half plane and statistics of the integral of a Brownian curve}, J. Phys. A {\bf 26}, pp. L1157-L1162.
\bibitem{sm} Schwarz, J. M. and Maimon, R. (2001) {\it First-passage-time exponent for higher-order random walks: Using L\'evy flights},  Phys. Rev. E {\bf 64} 016120 (pp. 10).
\bibitem{sa} Sparre-Andersen, E. (1953). {\it On the fluctuations of sums of random variables}, Math. Scand. {\bf 1} pp. 263-285; Math. Scand. {\bf 2} pp. 195-223.
\bibitem{ys} Sinai, Y. G. (1992). {\it Distribution of some functionals of the integral of a random walk}, Theor. Math. Phys. {\bf 90}, pp. 219-241. (Teor. Mat. Fiz. {\bf 90}, pp. 323-353).
\bibitem{twb08} Burkhardt, T. W. (2008). {\it First-passage and extreme-value statistics of a particle subject to a constant force plus a random force}, J. Stat. Phys. {\bf 133}, pp. 217-230.
\bibitem{twb00} Burkhardt, T. W. (2000). {\it Dynamics of absorption of a randomly accelerated particle}, J. Phys. A {\bf 33}, pp. L429-L432.
\bibitem{csb} Cornell, S. J., Swift, M. R. and Bray, A. J. (1998). {\it Inelastic collapse of a randomly forced particle}, Phys. Rev. Lett. {\bf 81}, pp. 1142-1145.
\bibitem{bfg} Burkhardt T. W., Franklin, J. and Gawronski, R. R. (2000). {\it Statistics of a confined, randomly accelerated particle with inelastic boundary conditions}, Phys. Rev. E. {\bf 61}, pp. 2376-2381.
\bibitem{bk1} Burkhardt, T. W. and Kotsev, S. N. (2004) {\it Equilibrium of a confined, randomly accelerated, inelastic particle: Is there inelastic collapse?}, Phys. Rev. E {\bf 70}, 026105 (6 pp.).
\bibitem{kb} Kotsev, S. N. and Burkhardt, T. W. (2005). {\it Randomly accelerated particle in a box: Mean absorption time for partially absorbing and inelastic boundaries}, Phys. Rev. E {\bf 71}, 046115 (7 pp.).
\bibitem{bk2} Burkhardt, T. W. and Kotsev, S. N. (2006). {\it Equilibrium statistics of an inelastically bouncing ball, subject to gravity and a random force}, Phys. Rev. E {\bf 73}, 046121 (7 pp.).
\bibitem{rk} Redner, S. and Krapivsky, P. L. (1996). {\it Diffusive escape in a nonlinear shear flow: Life and death at the edge of a windy cliff}, J. Stat. Phys. {\bf 82}, pp. 999-1014.
\bibitem{bg} Bray, A. J. and Gonos, P. (2005). {\it Survival of a diffusing particle in a transverse flow field}, J. Phys. A. {\bf 38}, pp. 5617-5626.
\bibitem{bm} Bray, A. J. and Majumdar, S. N. (2006). {\it Partial survival and crossing statistics for a diffusing particle in a transverse shear flow}, J. Phys. A. {\bf 39}, L625-L631.
\bibitem{fr} Franklin, J. N. and Rodemich, E. R. (1968). {\it Numerical analysis of an elliptic-parabolic partial diffeerential equation}, SIAM J. Numer. Anal. {\bf 4}, pp. 680-716.
\bibitem{mp} Masoliver, J. and Porr\`a, J. M. (1995). {\it Exact solution to the mean exit time problem for free inertial processes driven by Gaussian white noise}, Phys. Rev. Lett. {\bf 75}, pp. 189-192.
\bibitem{bb1} Bicout, D. J. and Burkhardt, T. W. (2000). {\it Absorption of a randomly accelerated particle: gambler's ruin in a different game}, J. Phys. A {\bf 33}, pp. 6835-6841.
\bibitem{twb97} Burkhardt, T. W. (1997). {\it Free energy of a semiflexible polymer in a tube and statistics of a randomly accelerated particle}, J. Phys. A {\bf 30}, pp. L167-L172.
\bibitem{ejg} Gumbel, E. J. (1958). {\it Statistics of Extremes}, Columbia University.
\bibitem{jg} Galambos, J. (1978). {\it The Asymptotic Theory of Extreme Order Statistics}, Wiley.
\bibitem{sc} Coles, S. (2001). {\it An Introduction to Statistical Modeling of Extreme
Values}, Springer.
\bibitem{chbs} Castillo. E., Hadi, A. S., Balakrishnan, N. and Sarabia, J. M. (2005). {\it Extreme Value and Related Models with Applications in Engineering and Science}, Wiley.
\bibitem{twbbudapest} Burkhardt, T. W., Gy\"orgyi, G., Moloney, N. R. and R\'acz, Z.
(2007) {\it Extreme statistics for time series: Distribution of the maximum relative to the initial value}, Phys. Rev. E {\bf 76}, 041119 (12 pp.).
\bibitem{rmr} Reymbaut, S., Majumdar, S. N. and Rosso, A. (2011). {\it The convex hull for a random acceleration process in two dimensions}, J. Phys. A {\bf 44}, 415001 (15 pp.).
\bibitem{bb2} Bicout, D. J. and Burkhardt, T. W. (2001). {\it Simulation of a semiflexible polymer in a narrow cylindrical pore}, J. Phys. A {\bf 34}, pp. 5745-5750.
\bibitem{ybg} Yang, Y., Burkhardt, T. W. and Gompper, G. (2007). {\it Free energy and extension of a semiflexible polymer in cylindrical confining geometries}, Phys. Rev. E {\bf 76}, 011804 (7 pp.).
\bibitem{byg} Burkhardt, T. W., Yang, Y. and Gompper, G. (2010). {\it Fluctuations of a long, semiflexible polymer in a narrow channel}, Phys. Rev. E {\bf 82}, 041801 (9 pp.).
\bibitem{to} Odijk, T. (1983) {\it On the statics and dynamics of confined or entangled stiff polymers}, Macromolecules {\bf 16}, pp. 1340-1344; (1986).  {\it Theory of lyotropic polymer liquid crystals}, Macromolecules {\bf 19}, pp. 2313-2329.
\bibitem{retal} Reisner, W., Morton, K. J., Riehn, R., Wang, Y. M., Yu, Z., Rosen, M.,
Sturm, J. C., Chou, S. Y., Frey, E. and Austin, R. H. (2005). {\it Statics and dynamics of single DNA molecules confined in nanochannels}, Phys. Rev. Lett. {\bf 94}, 196101 (4 pp.).
\bibitem{kp1} K\"oster, S., Steinhauser, D. and Pfohl, T. (2005). {\it Brownian motion of actin filaments in confining microchannels}, J. Phys. Condens. Matter {\bf 17}, S4091-S4104.
\bibitem{kp2} K\"oster, s., Kierfeld, J and Pfohl, T. (2008). {\it Characterization of single semiflexible filaments under geometric constraints}, Eur. Phys. J. E {\bf 25}, pp. 439-449.
\bibitem{kp3} K\"oster S. and Pfohl, T. (2009). {\it An in vitro model system for cytoskeletal confinement}, Cell Motility and the Cytoskeleton {\bf 66}, pp. 771-776.
\bibitem{mb} Majumdar, S. N. and Bray, A. J. (2001). {\it Spatial persistence of fluctuating interfaces}, Phys. Rev. Lett. {\bf 86}, pp. 3700-3703.
\bibitem{gbr} Golubovi\'c, L. and Bruinsma, R. (1991). {\it Surface diffusion and fluctuations of growing interfaces}, Phys. Rev. Lett. {\bf 66}, pp. 321-324.
\bibitem{dst} Das Sarma, S. and Tamborenea P. (1991). {\it A new universality class for kinetic growth: One-dimensional molecular-beam epitaxy}, Phys. Rev. Lett. {\bf 66}, pp. 325-328.
\bibitem{mhl} Maggs, A. C., Huse, D. A. and Leibler S. (1989) {\it Unbinding transitions of semi-flexible polymers}, Europhys. Lett. {\bf 8}, pp. 615-620.
\bibitem{gbu} Gompper, G. and Burkhardt, T. W. (1989) {\it Unbinding transition of semiflexible membranes in $(1+1)$ dimensions}, Phys. Rev. A, {\bf 40}, 6124-6127.
\bibitem{bll} Bundschuh, R., L\"assig, M. and Lipowsky, R. (2000). {\it Semi-flexible polymers with attractive interactions}, Eur. Phys. J. E {\bf 3}, pp. 295-306.
\bibitem{bl} Bundschuh, R. and L\"assig, M. (2002). {\it Delocalization transitions of semiflexible manifolds}, Phys. Rev. E {\bf 65}, 061502 (7 pp.).
\bibitem{bf} Benetatos, P. and Frey, E. (2003). {\it Depinning of semiflexible polymers}, Phys. Rev. E {\bf 67}, pp. 051108 (7 pp.).
\bibitem{cd} Caravenna, F. and Deuschel, J.-D. (2008). {\it Pinning and wetting transition for $(1+1)$-dimensional fields with Laplacian interaction}, Ann. Probab. {\bf 36}, pp. 2388-2433.
\bibitem{ee} Eisenriegler, E. (1993) {\it Polymers near Surfaces}, World Scientific.
\bibitem{pv} Valageas, P. (2009). {\it Statistical properties of the Burgers equation with Brownian initial velocity}, J. Stat. Phys. {\bf 134}, pp. 589-640.
\bibitem{hcs} Hilhorst, H. J., Calka, P. and Schehr, G. (2008). {\it Sylvester's question and the random acceleration process}, J. Stat. Mech. P10010 (25 pp.).
\bibitem{mrz} Majumdar, S. N., Rosso, A. and Zoia, A. (2010). {\it Time at which the maximum of a random acceleration process is reached}, J. Phys. A {\bf 43} 115001 (16 pp.).
\bibitem{levy} L\'evy, P. (1939). {\it Sur certains processus stochastiques homog\`enes}, Comp. Math. {\bf 7}, pp. 283-339.
\bibitem{wf} Feller W. (1968). {\it An Introduction to Probability Theory and Its Applications}, Wiley.
\end{thebibliography}
\end{document}